% Begining of Plain Tex version of the paper
%
\pageno=1
\footline={\hfil\folio\hfil} 
\baselineskip = 2\baselineskip
\centerline {DISSOCIATIVE AUTOIONIZATION IN $(1+2)$-PHOTON ABOVE THRESHOLD} 
\centerline {EXCITATION OF $H_2$ MOLECULES}
\vskip 2pc
\centerline {Krishna Rai Dastidar \footnote * {Electronic address : 
spkrd@iacs.ernet.in} and Ratan Kumar Das} 
\centerline {Department of Spectroscopy}
\centerline {Indian Association for the Cultivation of Science}
\centerline {Calcutta - 700032,\quad India} \vskip 2pc
\noindent {\bf Abstract.} We have theoretically studied the effect of 
dissociative autoionization on the photoelectron energy spectrum in 
$(1+2)$-photon above threshold ionization  (ATI) of $H_2$ molecules. 
We have considered excitation from the  ground $X^1\Sigma_g^+ (v=0,j)$ state 
to the doubly excited autoionizing states of $^1\Sigma_u^+$ and $^1\Pi_u$ 
symmetry, via the intermediate resonant $B^1\Sigma_u^+ (v=5,j)$ states. 
We have shown that the photoelectron energy spectrum is oscillatory in nature 
and shows three distinct peaks above the photoelectron energy $0.7$ ev. 
This feature has been observed in a recent experiment by Rottke et. al., 
J. Phys. B {\bf 30}, $4049 (1997)$. 
\vfill\eject
\noindent {\bf I. Introduction} \vskip 1pc

\noindent Studies on dissociative photoionization in molecular hydrogen
is being persued for several years, almost two decades, both experimentally 
and theoretically. In this process the molecule is excited above the 
dissociation threshold of the molecular ion by the single or multiphoton
absorption and the doubly excited autoionizing states play a significant role
to determine the photoelectron and the atomic ion spectrum. In most of the 
single photon experiments (Chung et al 1993, Dehmer and Dill 1978, Gardner and 
Samson 1975, Glass-Maujean 1986, Glass-Maujean et al 1995, He et al 1995, 
Ito et al 1988, 1996, Latimer et al 1992, 1993, Strathdee and Browning 1976) 
the molecule is excited from the ground ro-vibrational level of the ground 
electronic $X^1\Sigma_g^+$ state to the dissociation continuum above the 
ionization threshold leaving the signature of autoionization on the 
dissociative ionization yields. In most of the multiphoton experiments 
(Anderson et al 1984, Cornaggia et al 1986, Normand et al 1986, Pratt et al, 
1986, 1987,   Verschuur et al 1988, 1989, Xu et al 1989), the molecule is 
excited from the ground state to just above the ionization continuum 
(and below the dissociation continuum) or above the dissociation continuum, 
via  intermediate resonant rovibrational level of $B^1\Sigma_u^+$, 
$E,F^1\Sigma_g^+$ or $C^1\Pi_u$ states. In the above experiments, when the 
molecule is excited below the dissociation threshold (i.e. when only the bound 
molecular ions are produced in different rovibrational levels), the role of  
doubly excited states on rovibrational branching of the molecular ion has been 
studied; also, when the molecule is excited above the dissociation threshold 
(i.e. when  atomic ions and neutrals are produced as dissociative ionization 
products together with the bound molecular ions), the effect of autoionization 
via doubly excited states on the atomic ion spectrum and the photoelectron 
spectrum (PES), have been studied. Several theoretical studies ( Chupka 1987, 
Ganguly et al 1986, Ganguly and Rai Dastidar 1988, Hazi 1974, Hickman 1987, 
Kanfer and Shapiro 1983, Kirby et al 1979, 1981, Rai Dastidar et al 1986) have 
revealed that autoionization via doubly excited autoionizing (AI)
states play a dominant role in determining the autoionization lineshape and 
the spectrum of ionization yields i.e. photoelectrons and photoions. It has 
been shown that the interference of autoionization with ionization 
(Fano 1961), leaves its signature on (i) the autoionization lineshape, 
(ii) the rovibrational branching of the molecular ion in the ground state   
 and (iii) the spectrum of the dissociative ionization products (i.e. atomic 
ions and photoelectrons), for the  excitation  above the dissociation 
threshold of the molecular ion. 

For the single photon excitation from the ground state, only the 
{\it ungerade} states can be accessed and the lowest of these series is 
situated approximately $25.5$ ev above the ground state. In this case the 
Franck-condon transition  occurs at very low intenuclear distances 
(around $R=1.4 a.u.$) and  at very high continuum energy. But in case of 
resonance enhanced (m+n)-photon transition, the molecule is selectively 
excited to an intermediate rovibrational level of a Rydberg state and hence 
the Franck-condon trasition 
to the autoionizing state occurs at large internuclear distance. Therefore,
in this REMPI ( resonance enhanced multiphoton ionization ) technique, the
initial condition for dissociative ionization can be changed and the molecule
can be excited to different continuum energies, much  below the energy, 
reached in the single photon transition. Recently, experimental study (Rottke 
et al 1997) on $(3+2)$-photon dissociative ionization of $H_2$ molecules has 
been done by  selectively exciting the $B^1\Sigma_u^+(v=5,j)$ levels, from the 
ground $X^1\Sigma_g^+(v=0,j)$ states by three photon absorption and then to 
the dissociative continuum by two-photon
absorption from these intermediate resonant levels of $B^1\Sigma_u^+(v=5,j)$ 
state. Angle resolved photoelectron spectrum (PES) and ion spectrum have been 
studied and it has been found that the PES gives oscillatory structure with 
increase in energy. 

 In the present study, an 
attempt has been made to explain theoretically the origin of such oscillatory
behaviour of PES. In this calculation we have considered $(1+2)$-photon 
dissociative ionization of $H_2$ molecules. The transition scheme considered 
here is shown schematically in Fig. 1. By the single 
photon transition, 
the molecule is first excited resonantly to the $B^1\Sigma_u^+(v=5,j)$ state 
from the ground $X^1\Sigma_g^+(v=0,j)$ state of the $H_2$ molecule and by the 
subsequent two-photon transition the molecule is excited from this intermediate
 rovibrational level of $B^1\Sigma_u^+$ state
 to the dissociative continuum of the molecular ion in the 
ground state $(X^2\Sigma_g^+)$. In this transition,  the doubly excited states 
of {\it ungerade} symmetry are excited and hence the dissociative 
autoionization process via these AI states contributes to the PES. Our aim 
here is to show that the dissociative autoionization process can give rise to 
oscillatory photoelectron spectrum, as observed in the recent experiment 
(Rottke et al 1997).
\vskip 2pc
\noindent {\bf II. Transition Schematics} \vskip 1pc

\noindent In the experiment, the $H_2$ molecule is first excited from the 
ground $X^1\Sigma_g^+(v=0)$ state to the $B^1\Sigma_u^+ (v=5)$ state by
three photon absorption considering three transition schemes:
$$(i)P(1): X^1\Sigma_g^+(v=0,j=1) \to B^1\Sigma_u^+(v=5,j=0)$$
$$(ii)R(0): X^1\Sigma_g^+(v=0,j=0) \to B^1\Sigma_u^+(v=5,j=1)$$
$$(iii)R(2): X^1\Sigma_g^+(v=0,j=2) \to B^1\Sigma_u^+(v=5,j=3)$$
and then by two photon transition to the dissociative 
continuum and autoionizing states of {\ ungerade} symmetry. In our calculation
 we have considered (Fig. 1) single photon transition from the ground 
$X^1\Sigma_g^+(v=0,j=0,1,2)$ levels to the different rotational  levels of 
$B^1\Sigma_u^+(v=5)$ state by three transition schemes mentioned above and then
two photon transition from $B^1\Sigma_u^+(v=5)$ state with rotational
quantum numbers $J=0$,$1$ and $3$ to the continuum above the dissociation 
threshold.
In this process, excitation to four autoionizing states, two lowest
AI states of $^1\Sigma_u^+$ symmetry and the two lowest AI states of $^1\Pi_u$ 
symmetry in the $Q_1$ series has been considered. 
 To get the photoelectron spectrum,
 the autoionization decay from these AI states to the different
 dissociation continuua of the ground $X^2\Sigma_g^+$ state of the hydrogen 
molecular ion has been studied.
\vskip 2pc
\noindent {\bf III. Theory} \vskip 1pc

\noindent We have used the resolvent operator technique to obtain the 
dissociative
ionization and autoionization rate. In this technique a set of equations for
the matrix elements of resolvent operator ${G(z)}={1\over{z-H}}$, with respect 
to product states, such as $\vert g \rangle \vert n \rangle$, $\vert i \rangle 
\vert n-1 \rangle$, $\vert a \rangle \vert n-2 \rangle$, $\vert c_1 \rangle 
\vert n-2 \rangle$, $\vert b \rangle \vert n-3 \rangle$ and $\vert c_2 \rangle
\vert n-3 \rangle$. Here,
$\vert g \rangle$ and  $\vert i \rangle$ are the ground and the intermediate
resonant states of the bare molecule, $\vert a \rangle$ and $\vert b \rangle$ 
are 
the AI states of different symmetry excited by the single photon and two photon
absorption from the intermediate state of the molecule respectively and the
continuua $\vert c_1 \rangle$ and $\vert c_2 \rangle$ are  excited by the 
single
photon and two photon absorption from the intermediate state respectively (see 
Fig. 1). 
Hence the continuua are of different symmetry. The continuum 
$\vert c_2 \rangle$ will be of the same symmetry as the intermediate resonant 
state $\vert i \rangle$ and the continuum $\vert c_1 \rangle$ will be of 
opposite symmetry.
The $\vert n \rangle$ 's are the photon number states. H is the total 
hamiltonian
of the molecule+photon system. z is the complex energy of the system and the
imaginary part of z gives the decay rate of the system. 
For the $(1+2)$-photon
transitions considered here, the equations for the matrix elements of 
resolvent operator are given as follows:
$$(z-E_g)G_{gg} - D_{gi}G_{ig} = 1 \eqno(1)$$
$$(z-E_i)G_{ig} - D_{ia}G_{ag} - \int D_{ic_1}G_{c_1g} dE_{c_1} - D_{ig}G_{gg} 
= 0 \eqno(2)$$
$$(z-E_a)G_{ag} - \int V_{ac_1}G_{c_1g} dE_{c_1} - D_{ai}G_{ig} - 
\sum_b D_{ab}G_{bg} - \int D_{ac_2}G_{c_2g} dE_{c_2} = 0 \eqno(3)$$
$$(z-E_b)G_{bg} - D_{ba}G_{ag} - \int D_{bc_1}G_{c_1g} dE_{c_1}     
- \int V_{bc_2}G_{c_2g} dE_{c_2} = 0 \eqno(4)$$
$$(z-E_{c_1})G_{c_1g} - D_{c_1i}G_{ig} - V_{c_1a}G_{ag} - \sum_b D_{c_1b}
G_{bg} - \int D_{c_1c_2}G_{c_2g} dE_{c_2} = 0 \eqno(5)$$
$$(z-E_{c_2})G_{c_2g} - D_{c_2a}G_{ag} - \int D_{c_2c_1}G_{c_1g}dE_{c_1}
- \sum_b V_{c_2b}G_{bg} = 0 \eqno(6)$$
where $E_g = \epsilon_g + n\hbar \omega$, $E_i = \epsilon_i + (n-1)\hbar 
\omega$, $E_a = \epsilon_a + (n-2)\hbar \omega$, $E_b = \epsilon_b + 
(n-3)\hbar \omega$, $E_{c_1} = \epsilon_{c_1} + (n-2)\hbar \omega$ and 
$E_{c_2} = \epsilon_{c_2} + (n-3)\hbar \omega$ are the energies for 
the product states, i.e. the ground,
intermediate, autoionizing states and the two continuua respectively. Here 
$\epsilon$'s are the energies for the bare molecular states. $D_{jk}$'s are 
the dipole transition moments between the product states j and k. The 
summation over b indicates that contribution from all the autoionizing states 
excited
by this transition schemes has been considered. Formally obtaining $G_{c_1g}$
and $G_{c_2g}$ from equations (5) and (6) and substituting in equations (2),(3)
and (4) one can derive equations as follows:
$$Z_gG_{gg} - D_{gi}G_{ig} = 1 \eqno(7)$$
$$Z_iG_{ig} - D_{ig}G_{gg} - B_{ia}G_{ag} - \sum_b C_{ib}G_{bg} = 0 \eqno(8)$$
$$Z_aG_{ag} - B_{ai}G_{ig} - \sum_b K_{ab}G_{bg} = 0 \eqno(9)$$
$$Z_bG_{bg} - K_{ba}G_{ag} - C_{bi}G_{ig} = 0 \eqno(10)$$
where,
$$Z_g = z - E_g$$  $$Z_i = z - E_i - s_i + {i\over 2} \gamma_i$$ 
$$Z_a = z - E_a - S_a + {i \over 2} \Gamma_a - s_a + {i\over 2} \gamma_a + 
2R_{ac_1c_2a}$$  
$$Z_b = z - E_b - S_b + {i\over 2} \Gamma_b - s_b + 
{i\over 2} \gamma_b +2R_{bc_2c_1b}$$  
$$B_{ia} = Q_{ia} - R_{ic_1c_2a}$$ 
$$C_{ib} = Q_{ib} - R_{ic_1c_2b}$$  
$$K_{ab} = Q_{ab} - R_{ac_1c_2b} - R_{ac_2c_1b}$$
$$Q_{ia} = D_{ia} + {1\over A} \int {{D_{ic_1}V_{c_1a}}
\over z - E_{c_1}} dE_{c_1}$$  
$$Q_{ib} = {1\over A} \int {{D_{ic_1}D_{c_1b}}
\over z - E_{c_1}} dE_{c_1}$$  
$$ Q_{ab} = D_{ab} + {1\over A} 
\int{{D_{bc_1}V_{c_1a}} \over z - E_{c_1}} dE_{c_1} + {1\over A} \int
{{D_{ac_2}V_{c_2b}} \over z - E_{c_2}} dE_{c_2}$$ and
$$R_{jc_pc_qk} = {\pi^2 \over A} {\vert Y_{jc_p}D_{c_pc_q}X_{c_qk} \vert }$$
j,k corresponds to bound product states;  $c_{p,q}$ corresponds to product 
states for continuua;
$X_{c_qk}$ and $Y_{jc_p}$ correspond to coupling terms between the continuum 
$c_{p,q}$ and the bound 
states designated by k and j respectively. Here $A = {1 \over {1 + 
\pi^2 \vert D{c_1c_2} \vert^2}}$ ; 
$\gamma_j$ and $s_j$ are the photoionization width and ac-Stark shift,  
divided by the factor A,
of the bound states designated by j respectively; $\Gamma_{a,b}$ and 
$S_{a,b}$ are the 
autoionization width and the shift, divided by the factor A, of the 
autoionizing states designated by 
a and b respectively, due to the configuration interaction couplings 
$V_{ac_1}$ and $V_{bc_2}$ with  
the continuum adjacent to the AI states. The terms involving
integration over continuum (as in $Q_{ia}$, $Q_{ib}$, $Q_{ab}$ etc.) can be 
expressed as the sum of the principal value part and the imaginary part in the 
usual manner. In the present work we have neglected the principal value part
and used the pole approximation to derive the final formula for autoionization 
rate. Equations $(8) \to (10)$ can be 
rewritten in terms of the detunings ($\delta_i$, $\delta_a$ and $\delta_b$) 
from the intermediate and two AI states respectively by substituting 
$z = Z_g + E_g$ in the above equations. By solving above equations one can 
formally obtain the matrix elements for the resolvent operator and by inverse 
Laplace transformation of these matrix elements one can get corresponding 
matrix elements for the evolution operator. Hence the probability for 
ionization
can be written as: $P(t) = 1 - \vert U_{gg}(t) \vert^2 - \vert U_{ig}(t)
\vert^2 - \vert U_{ag}(t) \vert^2 - \vert U_{bg}(t) \vert^2$.

In the weak field limit i.e. considering dipole transition strength from the 
ground state to the intermediate state to be much weaker than that for the 
ionizing transitions and the autoionizing decay, one can write down 
$G_{gg} = {1 \over {Z_g - Z_0}}$ , where  $$Z_0 = {\vert D_{ig} \vert^2 \over 
{ Z_i - \sum_b {C_{bi}^2 \over Z_b} - {{\left [B_{ia} + \sum_b {{K_{ab}C_{bi}}
\over Z_b}\right ]^2}\over {Z_a - \sum_b {{K_{ab}}^2 \over Z_b}}}}}$$
and hence the rate for ionization  ${dP \over dt} = - 2 Im Z_0$.

By neglecting the higher order contributions, it can be shown that the rate 
for dissociative autoionization via the AI states $\vert b \rangle$  can be 
written as: 
$$ {dP_{2} \over dt} =
{4\pi^3{\vert D_{gi} \vert^2 \vert D_{ic_1}\vert^2 \vert D_{c_1b} \vert^2 
\vert V_{bc_2} \vert^2}
\over {CA^3 {\vert Z_b \vert^2}}} \eqno (11)$$ 
where $ C = \vert {Z_i - \sum_b {{C_{bi}}^2\over Z_b}} \vert^2 $ ; 
 $dP_2 \over dt$ can also be expressed as
$${dP_2\over dt} = {{\cal M}_{gc_2}}F \eqno (12)$$
where, ${F} = {4\pi^3\over {CA^3 {\vert Z_b \vert^2 }}}$, is the lineshape 
factor and the modified dissociative autoionization  matrix element square 
${{\cal M}_{gc_2}} = {\vert D_{gi} \vert^2 \vert D_{ic_1} \vert^2 D_{c_1b} 
\vert^2 \vert V_{bc_2} \vert^2}$
\vskip 2pc

\noindent {\bf IV. Calculations} \vskip 1pc

\noindent In the present study we have considered contribution from the 
dissociative 
autoionization via two lowest AI states of $^1\Sigma_u^+$ symmetry and two 
lowest AI states of $^1\Pi_u$ symmetry. The energies and autoionization widths 
for these four states are available in the literature {Guberman 1983, Kirby et 
al 1981, Tennyson 1996). As mentioned before, we have considered here 
(1+2)-photon transition from the ground state to the autoionizing states via 
the intermediate resonant $B^1\Sigma_u^+$ state. 
Therefore, to calculate the two-photon term via the continuum, one needs to 
have the electronic wavefunctions for the AI states. The wavefunctions for two 
lowest AI states of $^1\Sigma_u^+$ symmetry have been obtained by 
minimizing the energy variationally. The molecular basis set used 
is a multi-configuration minimal basis constructed with linear combination of 
Slater-type atomic orbitals. We have used seven configuration wavefunctions 
constructed with atomic orbitatls of s and p symmetry and the principal 
quantum number upto $4$. At each value of R (internuclear separation), 
all the exponents have been varied to obtain minimum energy. The energies 
obtained are in good agreement with the previous values (Guberman 1983) 
except at lower values of internuclear seperation $R \le 1.4 a.u.$. 
Calculated energies for the second $^1\Sigma_u^+$ AI state differ by  
$\le .006$ 
a.u. from the previous values (Guberman 1983), except at $R=1.0$ a.u., where 
the difference is .017 a.u.. For the first $^1\Sigma_u^+$ AI state, the 
difference in energies (our results and Guberman's results) are $\le .0086$ 
a.u., except at $R=1.0$ and $R=1.4$ a.u., where the differences are $.023$ 
a.u. and $.012$ a.u. respectively. The electronic wavefunctions for the ground
 and the intermediate states have been obtained from our previous calculations 
(Khatun and Rai Dastidar 1995). For the calculation of the complex nuclear 
wavefunctions using complex potential for the AI states, we have used modified 
Numerov method (Lane and Geltman 1967, Rai Dastidar and Rai Dastidar 1979). 

To calculate two-photon dipole transition moment between $B^1\Sigma_u^+$ state
and the AI states (i.e. to evaluate $Q_{ib}$; see section III) we have used the
pole approximation to integrate over the continuum energy ($E_{c_1}$), and 
hence considered contributions from the continuua associated with the 
bound vibrational levels of the $X^2\Sigma_g^+$ state of $H_2^+$ ion, which 
are energetically below the continuum energy excited by the first photon from 
the $B^1\Sigma_u^+$ state. We have used the multipolar gauge to evaluate dipole
transition moments, but one can also use the minimal coupling gauge.
 Since, in the present work, the two-photon term acts as a constant 
(with respect to PE energy) multiplying
factor to the autoionization amplitude-square (see equation 12), the structure 
of the PES is independent of the choice of gauge. However, the scaling factor 
for PES can be different,  depending on  the magnitude of two-photon 
term in two different gauges.  
For the present study we have calculated the two-photon dipole transition 
moments for the transition from the $B^1\Sigma_u^+$ state to the two lowest AI 
states of $^1\Sigma_u^+$ symmetry only. It is found that the  second AI state 
of $^1\Sigma_u^+$ symmetry is energetically very close to 
the lowest AI state of $^1\Pi_u$ symmetry (Guberman 1983, Tennyson 1996)) and 
the difference between the 
values of effective principal quantum numbers for these two states are also 
very small (Tennyson 1996). Therefore for the present calculation we have used 
the same 
two-photon dipole transition term for the transition to the lowest AI state 
of $^1\Pi_u$ symmetry as that calculated for the transition to the second AI 
state of $^1\Sigma_u^+$ symmetry. For the second $^1\Pi_u$ state we have used 
the same two-photon term and it is found that the contribution from this 
second $^1\Pi_u$ state is much smaller than that from the second 
$^1\Sigma_u^+$ state and the lowest $^1\Pi_u$ state. 

The electronic configuration interaction term between AI 
state and the continuum has been derived as $V_{bc_2}(R) = \sqrt{\Gamma_b(R) 
\over{2\pi}}$, where autoionization widths $\Gamma_b(R)$ have been obtained 
from the literature (Tennyson 1996) and this value of $V_{bc_2}$ has been 
considered to be arising from $l=1$ partial wave only. This is a valid 
approximation, since $l=1$ is the most dominant contribution for 
autoionization from $^1\Sigma_u^+$ AI states (Kirby et al 1981). To obtain 
autoionization amplitude, integration over nuclear coordinates have been done 
considering explicit dependence of the electronic transition amplitudes on 
the internuclear seperation. Similarly, to obtain dipole transition amplitudes,
 explicit dependence of electronic dipole transition moments on the 
internuclear separation (for bound-bound transition) and also on the 
photoelectron energy (for bound-continuum transition) has been considered for
nuclear integration. 

The autoionizing states lying above the dissociation continuum of the 
molecular ion can decay into the dissociation continuua lying between the final
 continuum energy excited by two-photon transition from the $B^1\Sigma_u^+$ 
state 
and the dissociation threshold of the ground state ($X^2\Sigma_g^+$) of the 
molecular ion. Hence  the total energy  $E_t$, 
can be expressed as $E_t = E_e + D_e$, where $E_e$ is the 
photoelectron energy and $D_e$ is the energy of the dissociation continuum of 
the molecular ion.
We have calculated the dissociative autoionization matrix element for the 
transitions to a range of 
dissociation continuum of the ground state of the $H_2^+$ ion, chosen 
(arbitrarily) to lie on a grid of $.01$ ev spacing.
\vskip 2pc 
\noindent {\bf V. Results and Discussions} \vskip 1pc

\noindent A schematic diagram for the transitions considered here is  shown 
in fig.1. For  different molecular potential energy curves we refer to Fig.1 
of Rottke et al, 1997. The two continuua shown in Fig. 1 are of different 
symmetry and correspond to the first ionization threshold ($H_2^+ 
(X^2\Sigma_g^+) + e$) of the neutral molecule.
Results for three different transition schemes ($P(1)$, $R(0)$ and $R(2)$),
i.e. for ionizing transitions from $J=0$, $J=1$ and $J=3$ of 
$B^1\Sigma_u^+(v=5)$ 
state are shown in Figs. 2,3 and 4 respectively.  In these figures we 
have plotted the
dissociative autoionization rate divided by $I^3F$ ( where I is 
the laser intensity and F is the line shape factor, see equ. 12), as a 
function of photoelectron energy. The factor $I^3F$, which is independent
of the photoelectron energy, merely plays the role of a scaling factor
as far as the shape and the structure of the PES is concerned. 
The experimental PES is given in arbitrary units and here we compare our 
theoretical PES (scaled by the factor $I^3F$) with experiment.
It is to be mentioned here that, for photoelectron energies less than $0.7$ ev,
 the contribution from the ionization of neutral atoms is dominating (Rottke 
et al $1997$) and hence the effect of dissociative autoionization will be 
prominent only for energies ${\ge 0.7}$ ev. Therefore, 
we will discuss the features obtained in PES for energies above $0.7$ ev 
(  experimental results also show PES above $0.7$ ev, see fig.2 of Rottke et 
al, 1997). 

For this calculation we have taken $A = 1$, since the value of 
$\vert D_{c_1c_2} \vert^2$ is much less than unity for this calculation. The 
continuum-continuum transition amplitude (${\vert D_{c_1c_2} \vert}$) will be 
significant only for photoelectron energies very close to each other (Gordon 
1929) and for the present study the photoelectron energies for the 
transition to the lower continuum ( due to the absorption of single photon 
from the $B^1\Sigma_u^+(v=5)$ state ) are less than $.02$ a.u. (i.e. less than 
$0.54$ ev ). Hence the value of $\vert D_{c_1c_2} \vert^2$ will be significant 
for photoelectron energies less than $0.54$ ev for continuum-continuum 
transition. To get an estimate, we have calculated electronic dipole 
transition moment for continuum-continuum transition (Gordon 1929) and the 
overlap between the three lowest vibrational levels of molecular ion in the 
ground state, with different dissociation continuum  of the same. It is 
found that the contribution from this term is orders of magnitude smaller than
 unity (Rai Dastidar 1996). 

In the figures ($2,3$ and $4$), individual contributions to dissociative 
autoionization from four ( first two $^1\Sigma_u^+$ and first two $^1\Pi_u$ )
autoionizing states have been shown, together with their sum. It is found 
that the PES is oscillatory in nature and the dissociative autoionization from 
the second AI state of $^1\Sigma_u^+$ symmetry is the most important 
contribution. We have checked that the oscillation is essentially the same as 
in the Franck-Condon overlap (mod-squared) between the complex nuclear 
wavefunction of the autoionizing state and the nuclear continuum of the 
molecular ion in the ground state, as a function of photoelectron energy. 
The PES for autoionization from three AI  ( second $^1\Sigma_u^+$ 
and the two $^1\Pi_u$) states show three peaks above the photoelectron energy 
greater than $ 0.7$ ev and the peak positions are almost the same for these 
three PES. But the contribution from the lowest $^1\Sigma_u^+$ state is much 
smaller than that from the second $^1\Sigma_u^+$ state and shows an out of 
phase oscillation. The presence of three peaks in PES, above $0.7$ ev of 
photoelectron energy, has also been experimentally observed (Rottke et al 
1997). From the figure for PES (Fig. 2 of Rottke et al 1997), the peak 
positions can be approximately calculated and we find that the peaks are
shited towards higher energy than that obtained in the present calculation. 
These shifts are different for three transition schemes and
the values are approximately within $0.23-0.24 $ ev, $0.17-0.2$ ev and 
$0.04-0.09 $ ev for the first, second and the third peak respectively. 
 It is to be mentioned here that the 
shifts of the autoionizing states due to autoionization decay have not been 
considered in our calculation. By using the approximate formula (Ganguly and 
Rai Dastidar 1988,  Olsen 1982)  ${S_b} = {-}{3\over {2\pi}} \Gamma_b$, one 
can calculate the approximate values of shift ($S_b$) for the second AI state 
of $^1\Sigma_u^+$ symmetry (using the values of width tabulated by Tennyson,
1996). Typical values of shift varies from $-0.02$ to $-0.16$ ev, in the range 
of internuclear separation, $R=3.0$ to $4.0$ and since this value is 
subtracted from the energy of the AI state, electronic potential energies will 
be shifted upwards. Hence the peak positions in figs. $2,3$ and $4$ will be 
shifted towards higher energy and this shifting will be different for different
photoelectron energies. The values of shifts are 0.2, 0.16 and 0.14 ev  for 
the first, second and the third peak respectively. It is found that, after 
shifting, the peak positions are almost the same as those observed in the 
experiment. It is within $.03$ to $.04$ ev,  $.01$ to $.04$ ev and $.05$ to 
$0.1$ ev for the first, second and the third peak respectively.
But for comparing our results with the experimental PES, the resolution factor 
for the experiment has to be taken into account (in particular for determining 
the peak positions). It has been mentioned by Rottke et al, 1997 that the 
resolution is $.08$ ev at $2$ ev photoelectron energy but that for energies 
lower than $2$ ev has not been mentioned. We expect that the resolution will 
be better as the photoelectron energy is decreased and we find that the 
agreement is better for the first two peak positions than the third one.  

By comparing figures $2, 3$ and $4$, it is found that the PES for three 
transition schemes are of the same order of magnitude and the PES for 
transition from $J=1$ level (Fig. 3) is greater in magnitude than that from 
$J=0$ (Fig. 2) and $J=3$ (Fig. 4) levels of $B^1\Sigma_u^+(v=5)$ state. 
However, in the experiment, although the contribution for these three 
transition schemes are of the same order (in arbitrary unit), photoelectron 
yield for transition from J=0 level of $B^1\Sigma_u^+(v=5)$ state is greater 
than that from other two transitions. By analyzing our results, we find that 
the (1+2)-photon term for transition to the second $^1\Sigma_u^+$ AI state is
 greater for excitation from $j=0$ level of the ground $X^1\Sigma_g^+(v=0)$ 
state   than that from other j levels. In the presnt work, we have done 
ab-initio calculations to obtain the  PES for dissociative autoionization from 
two $^1\Sigma_u^+$ AI states. But to obtain PES from two $^1\Pi_u$ AI states we
 have  used the same (1+2)-photon term as calculated for second $^1\Sigma_u^+$ 
state. Hence the scaling factor for contribution from these two AI states 
($^1\Pi_u$) may  become different for three transition schemes, if the actual 
values of (1+2)-photon term for excitation to these two $^1\Pi_u$ states are 
obtained. Moreover, the values of (1+2)-photon term may differ in two different
gauges (minimal coupling and multipolar).
Therefore, at this stage, this difference between our results and 
the experiment may arise, as far as the relative magnitudes of three PES (for 
transitions from $J=0,1$,and $3$) are concerned. Nevertheless, further study 
in this field is warranted.

In conclusion, we have shown that the oscillatory nature of  the 
experimentally observed PES in (3+2)-photon above threshold ionization of 
$H_2$ molecules, can be accounted for by considering the dissociative 
autoionization of the doubly excited autoionizing states (of {\it ungerade} 
symmetry) above the dissociative ionization threshold. 
\vskip 2pc
\noindent {\bf Acknowledgement} \vskip 1pc

\noindent This work has been sponsored by DST, Govt. of India, Project No. 
$SP/S2/L-09/94$. RKD is grateful to DST, Govt. of India for awarding Junior 
Research Fellowship. We are thankful to Prof. A. Suzor-Weiner for helpful 
discussions and drawing our attention to the paper by Tennyson, 1996.
We are also thankful to Prof. T.K. Rai Dastidar for critically reading the 
manuscript.
\vskip 2pc
\noindent {\bf References} \vskip 1pc
	
\noindent Anderson L, Kubiak G D, and Zare R N $1984$ {\it Chem. Phys. Lett.} 
{\bf 105} $22$

\noindent Chung Y M, Lee E M, Masuoka T and Samson J A R, $1993$ {\it J. Chem.
 Phys.}  {\bf 99} $885$

\noindent Chupka W A $1987$ {\it J. Chem. Phys.} {\bf 87} $1488$

\noindent Cornaggia C, Normand D, Morellec J, Mainfray G and Manus C $1986$ 
{\it Phys. Rev. A} {\bf 34} $207$  

\noindent Dehmer J L, and Dill D $1978$ {\it Phys. Rev. A} {\bf 18} $164$

\noindent Fano U $1961$ {\it Phys. Rev.} {\bf 124} $1866$

\noindent Ganguly S, Rai Dastidar K  and Rai Dastidar T K $1986$ {\it Phys. 
Rev. A} {\bf 33} $337$

\noindent Ganguly S and Rai Dastidar K $1988$ {\it Phys. Rev. A } {\bf 37} 
$1363$

\noindent Gardner J L and Samson J A R $1975$ {\it Phys. Rev. A} {\bf 12} 
$1404$

\noindent Glass-Maujean M $1986$ {\it J. Chem. Phys.} {\bf 85} $4830$

\noindent Glass-Maujean M, Frohlich H and Martin P 1995 {\it Phys. Rev. A} 
{\bf 52} $4622$

\noindent Gordon W $1929$ {\it Anns. of Phys.} {\bf 5} $1031$

\noindent Guberman S L $1983$ {\it J. Chem. Phys.} {\bf 78} $1404$

\noindent Hazi A U $1974$ {\it J. Chem. Phys.} {\bf 60} $4358$

\noindent He Z X, Cutler J N, Southworth S H, Hughey L R and Samson J A R 
$1995$ {\it J. Chem. Phys.} {\bf 103} $3912$

\noindent Hickman A P $1987$ {\it Phys. Rev. Lett.} {\bf 59} $1553$

\noindent Ito K, Hall R I, and Ukai M $1996$ {\it J. Chem. Phys.} 
{\bf 104} $8449$

\noindent Ito K, Lablanquie P, Guyon P-M and Nenner I $1988$ 
{\it Chem. Phys. Lett.} {\bf 151} $121$

\noindent Kanfer S and Shapiro M $1983$ {\it J Phys. B: At. Mol. Phys.} 
{\bf 16} $L655$

\noindent Khatun J and Rai Dastidar K $1995$ {\it Phys. Rev. A} {\bf 52} $2971$

\noindent Kirby K, Guberman S and Dalgarno A $1979$ {\it J. Chem. Phys.} 
{\bf 70} $4635$

\noindent Kirby K, Uzer T, Allison A C and Dalgarno A $1981$ {\it J. Chem. 
Phys.} {\bf 75} $2820$

\noindent Lane N F and Geltman S $1967$ {\it Phys. Rev.} {\bf 160} $53$

\noindent Latimer C J, Dunn K F, Kouchi N, McDonald M A, Srigengan V and 
Geddes J $1993$ {\it J. Phys. B: At. Mol. Opt. Phys.} {\bf 26} $L595$

\noindent Latimer C J, Irvine A D, McDonald M A and Savage O G $1992$ 
{\it J. Phys. B: At. Mol. Opt. Phys.} {\bf 25} $L211$

\noindent Olsen T $1982$ Ph.D. Dissertation, University of Southern California

\noindent Normand D, Cornaggia C and Morellec J $1986$ {\it J. Phys. 
B: At. Mol. Phys.} {\bf 19} $2881$

\noindent Pratt S T, Dehmer P M and  Dehmer J L $1986$ {\it J. Chem. Phys.} 
{\bf 85} $3379$

\noindent ---- $1987$ {\it J. Chem. Phys.} {\bf 86} $1727$

\noindent Rai Dastidar K, Ganguly S and Rai Dastidar T K $1986$ 
{\it Phys. Rev. A} {\bf 33} $2106$

\noindent Rai Dastidar K $1996$ {\it Phys. Rev. A} {\bf 53} $2881$
\noindent Rai Dastidar K and Rai Dastidar T K $1979$ {\it J. Phys. Soc. of 
Japan} {\bf 46} $1288$

\noindent Rottke H, Ludwig J and Sandner W $1997$ {\it J. Phys. B: At. Mol. 
Opt. Phys.} {\bf 30} $4049$

\noindent Strathdee S and Browning R $1976$ {\it J. Phys. B: At. Mol. Phys.} 
{\bf 9} $L505$

\noindent -----$1979$ {\it J. Phys. B: At. Mol. Phys.} {\bf 12} $1789$

\noindent Tennyson J $1996$ {\it Atomic Data and Nuclear Data Tables} {\bf 64} 
$253$

\noindent Verschuur J W J, Noordan L D, Bonnie J H M and van Linden van den 
Heuvell H B $1988$ {\it Chem. Phys. Lett.} {\bf 146} $283$

\noindent Verschuur J W J and van Linden van den Heuvell H B $1989$ 
{\it Chem. Phys.} {\bf 129} $1$

\noindent Xu E, Hickman A P, Kachru R, Tsuboi T and H Helm $1989$ 
{\it Phys. Rev. A} {\bf 40} $7031$

\noindent Xu E Y, Tsuboi T, Kachru R and Helm H $1987$ {\it Phys. Rev. A} 
{\bf 36} $5645$
\vskip 2pc
\noindent {\bf Figure Captions} \vskip 1pc

\parindent=4em
\item {fig. 1:} Schematic diagram for $(1+2)$-photon above threshold 
ionization. Two continuua $\vert c_1 \rangle$ and $\vert c_2 \rangle$ are of 
{\it gerade} and {\it ungerade} symmetry  respectively and correspond to the 
first ionization threshold ($H_2^+(X^2\Sigma_g^+)+e$) of the hydrogen molecule.
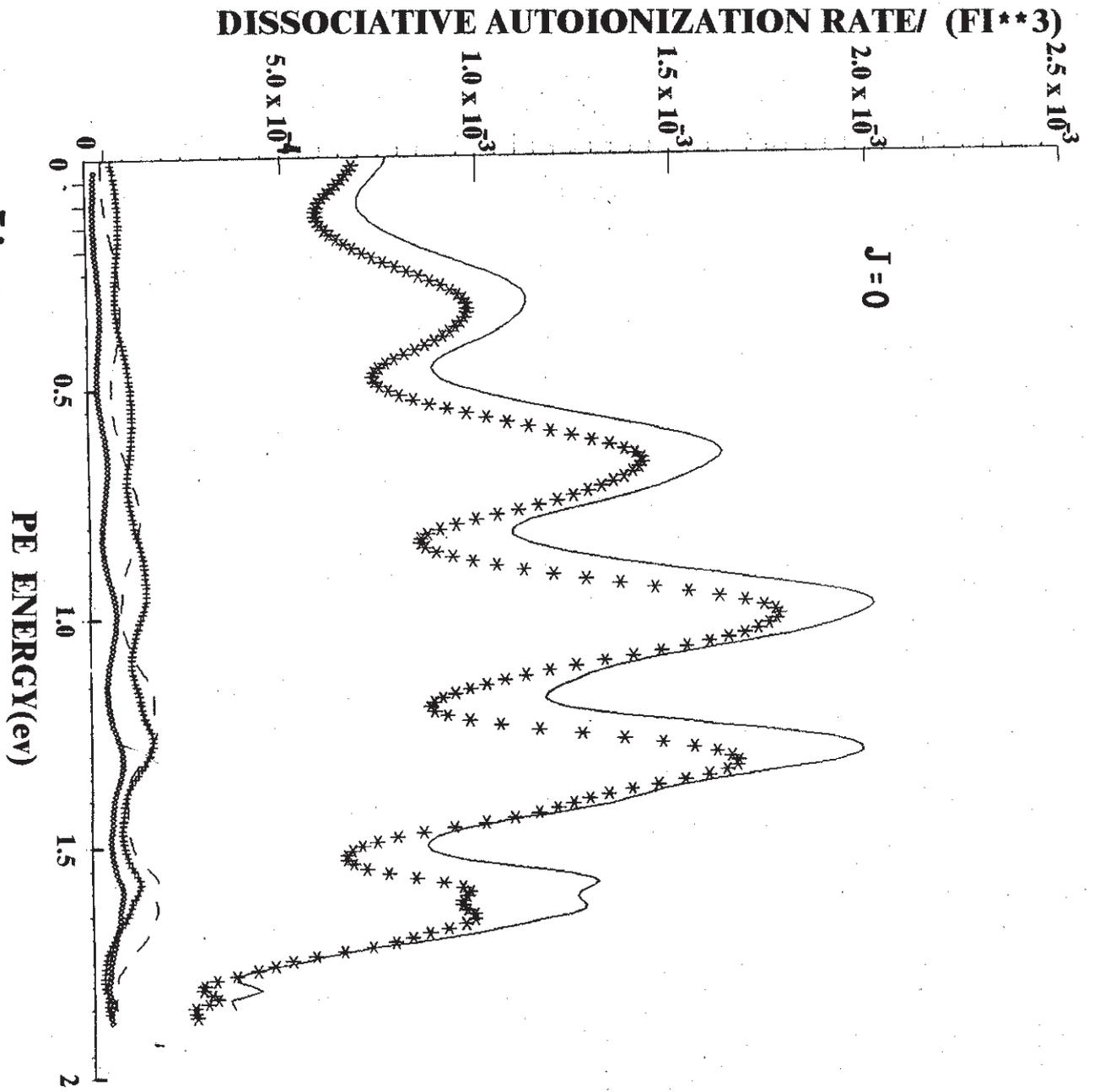
\item {fig. 2:} Dissociative autoionization rate divided by $I^3F$ (see text, 
equ. 12), where I is the laser intensity and F is the lineshape factor, has 
been plotted as a function of photoelectron energy,
for two-photo ionizing transition from rotational level, J=0 of 
$B^1\Sigma_u^+(v=5)$ state. Contributions from four AI states and their sum 
have been shown separately. Legends for the curves are: - - - contribution 
from the lowest $^1\Sigma_u^+$ AI state; **** same from second $^1\Sigma_u^+$ 
AI state; ++++ same from lowest $^1\Pi_u$ AI state; $\diamondsuit \diamondsuit 
\diamondsuit$ same from the second $^1\Pi_u$ state and \--- sum total of the 
four contributions.
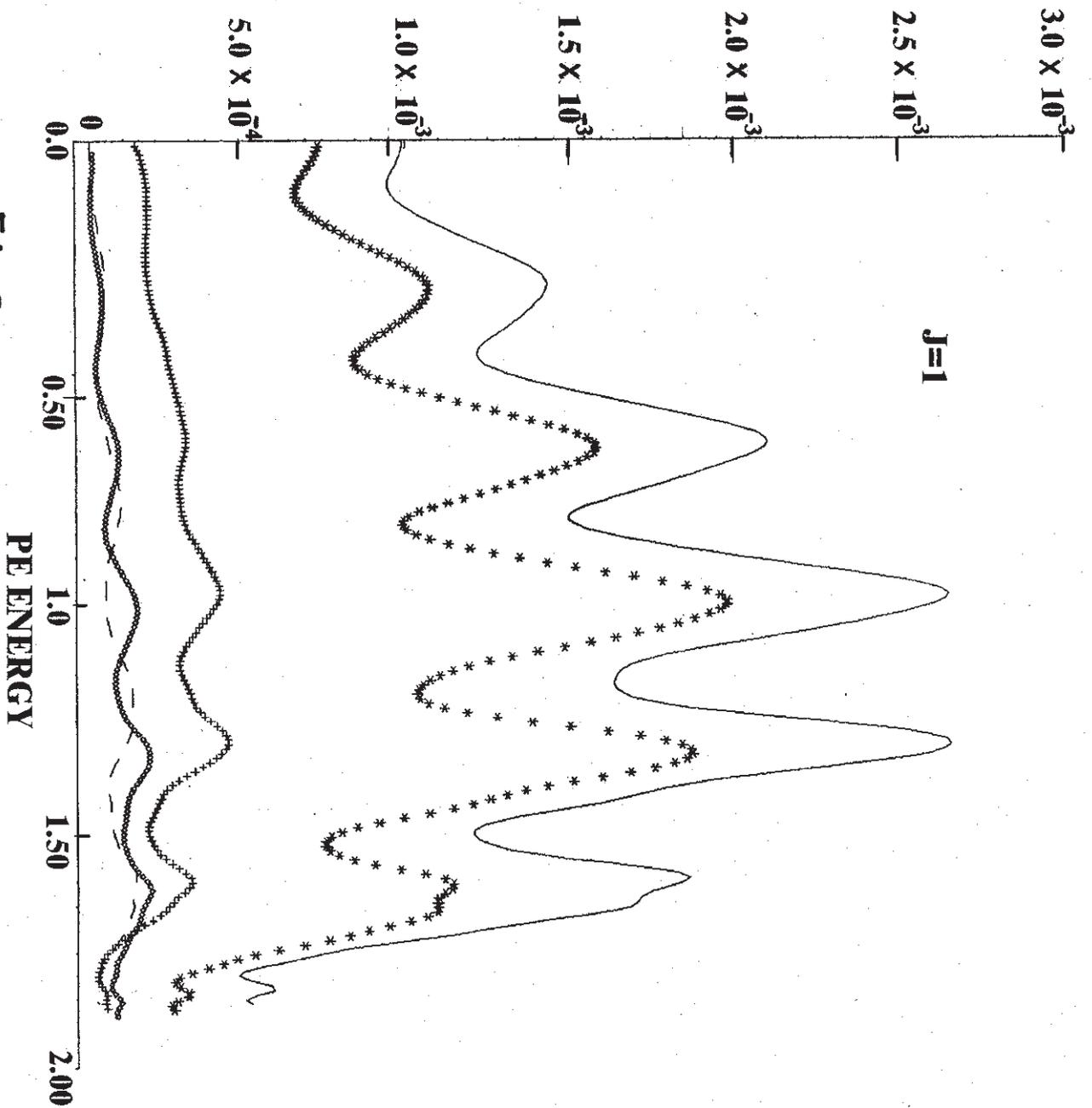
\item {fig. 3:} Same as in Fig. 2, for transition from J=1 level of 
$B^1\Sigma_u^+(v=5)$ state.
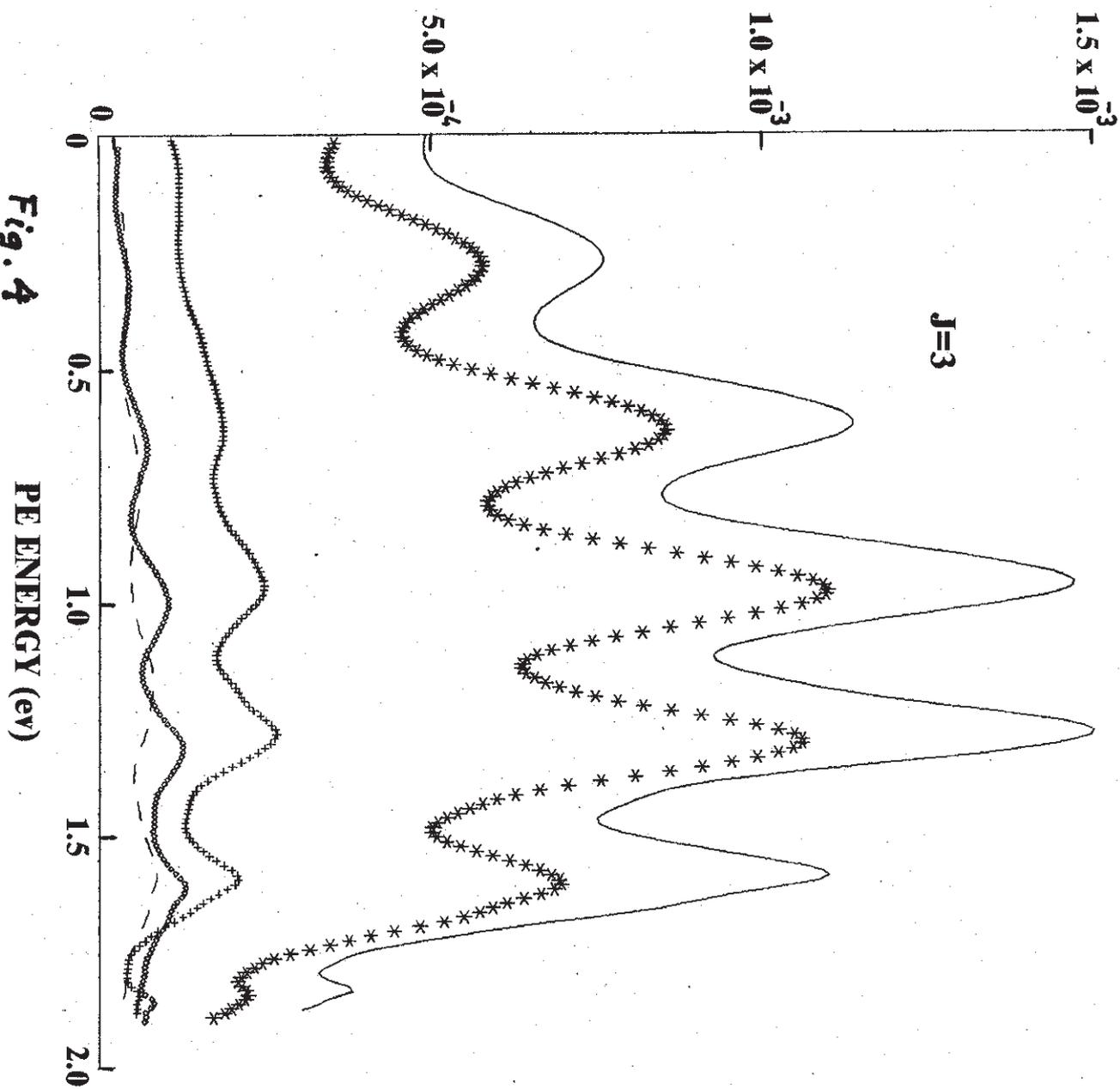
\item {fig. 4:} Same as in Fig. 2, for transition from J=3 level of 
$B^1\Sigma_u^+(v=5)$ state.
\bye